\documentclass[aip, amsmath,amssymb, reprint]{revtex4-1}
\usepackage{graphicx}
\usepackage{dcolumn}
\usepackage{bm}
\usepackage[version=4]{mhchem} 
\usepackage{hypernat}
\usepackage{hyperref}
\usepackage{xcolor, soul}
\makeatletter
\def\@email#1#2{%
 \endgroup
 \patchcmd{\titleblock@produce}
  {\frontmatter@RRAPformat}
  {\frontmatter@RRAPformat{\produce@RRAP{*#1\href{mailto:#2}{#2}}}\frontmatter@RRAPformat}
  {}{}
}%
\makeatother
\begin{document}

\preprint{AIP/123-QED}

\title{The study of electronic, structural, mechanical, and piezoelectric properties of bulk NbOX$_2$ (X = Cl, Br, and I) using density functional theory}
\author{L. D. Tamang}
\affiliation{Department of Physics, Mizoram University, Aizawl-796004, India}%
\affiliation{Physical Sciences Research Center (PSRC), Department of Physics, Pachhunga University College,  Aizawl-796001, India}
\author{B. Chettri}%
\affiliation{Department of Physics, Mizoram University, Aizawl-796004, India}%
\affiliation{Physical Sciences Research Center (PSRC), Department of Physics, Pachhunga University College,  Aizawl-796001, India}
\author{L. Celestine}
\affiliation{Department of Physics, Mizoram University, Aizawl-796004, India}%
\affiliation{Physical Sciences Research Center (PSRC), Department of Physics, Pachhunga University College,  Aizawl-796001, India}
\author{R. Zosiamliana}
\affiliation{Department of Physics, Mizoram University, Aizawl-796004, India}%
\affiliation{Physical Sciences Research Center (PSRC), Department of Physics, Pachhunga University College,  Aizawl-796001, India}
\author{S. Gurung}
 \affiliation{Physical Sciences Research Center (PSRC), Department of Physics, Pachhunga University College,  Aizawl-796001, India}
\author{A. Laref}
 \affiliation{Department of Physics and Astronomy, College of Science, King Saud University, Riyadh, 11451, Saudi Arabia}%

\author{Shalika R. Bhandari}
 \email{shalikram.bhandari@bmc.tu.edu.np}
\affiliation{Department of Physics, Bhairahawa Multiple Campus, Tribhuvan University, Siddarthanagar-32900, Rupandehi, Nepal}%

\author{Tatyana Orlova}
 \email{ayntraxt@gmail.com}
\affiliation{Department of Physics and Its Teaching Methods, Tashkent State Pedagogical University, Tashkent, Uzbekistan}

\author{D. P. Rai}
 \email{dibyaprakashrai@gmail.com}
\affiliation{Department of Physics, Mizoram University, Aizawl-796004, India}%
\date{\today}

\begin{abstract}
In this work, we have performed a comprehensive study of dielectric materials NbOX$_2$(X= Cl, Br, and I) within the framework of density functional theory, incorporating both conventional and hybrid functionals. Our studies focuses on the structure, electronic, elastic and piezoelectric properties. Piezoelectricity is an innovative avenue to extract energy by manipulating the material's structures via mechanical stress. The use of non-lead based material for piezoelectricity added an advantage of greener approach. Among the investigated materials, bulk NbOI$_2$ exhibits the highest piezoelectric response of 6.32 Cm$^{-2}$, which is $\sim 31\%$ higher than NbOCl$_2$, NbOBr$_2$ and even the lead zirconate titanate. 
\end{abstract}

\maketitle

\section{\label{sec:level1}Introduction }

The burning of fossil fuels and their harmful byproducts are the root causes for global warming and adverse effects on the environment, leading to natural disasters. Environmental pollution is not only a current concern for life sustainability, but energy shortage is also another alarming factor. To meet the current energy crisis, scientists and researchers are looking for alternative ways to generate energy in a greener way with zero emission of carbon.\cite{houghton2001science,soeder2021fossil} 

Nowadays, energy production via eco-friendly and cost-effective sources is the key research interest.\cite{brown2018response} There are various renewable energy sources like solar cells,\cite{al2022photovoltaic} hydropower,\cite{yuksel2010hydropower} geothermal,\cite{barbier2002geothermal} wind power,\cite{csahin2004progress} etc. These renewable approaches are highly innovative and much cleaner however, there are some limitations in terms of portability, mobility, and compatibility because of their huge size. Other drawbacks include the use of toxic elements such as lead (PbI$_2$), tin (SnI$_2$), cadmium, silicon, and copper in the components of solar cells, which are harmful to both the ecosystem and human health.\cite{kwak2020potential} In the case of wind energy, it lacks proper storage processes that hinder its full potential for energy harvest, therefore it is only able to produce 10\% of the current energy demand.
To overcome these problems, researchers are focusing mainly on harvesting kinetic energy (such as motion, vibrations, or shaking) because of their perennial availability, which is based on three transducer mechanisms: piezoelectric, electromagnetic, and triboelectric.
Among these processes, piezoelectricity is mostly prevalent innovative techniques to extract energy from the manipulation of crystal structure via applied mechanical stress, which is environmentally benign.\cite{hao2019progress,bowen2014piezoelectric,lalengmawia2024comprehensive} Despite its superior power densities, high energy conversion efficiency, and high scalability, the mass-scale application for energy generation is still questionable.\cite{bowen2014piezoelectric,li2014energy} So far piezoelectric applications are extensively integrated into various fields such as sensors,\cite{tressler1998piezoelectric,lang2006review,khan2015flexible} actuators,\cite{wang2010lateral,polla1998processing} nanogenerators,\cite{hu2015recent,tamang2015dna} portable electronic gadgets,\cite{mitcheson2008energy,xie2015high} and biomedicinal field.\cite{surmenev2019hybrid,motamedi2017piezoelectric}
The piezoelectric phenomenon is mainly exhibited by the ferroelectric materials having a wide band gap with broken inversion symmetry (i.e, non-centrosymmetric).\cite{kumar2015energy} This phenomenon was first discovered by Pierre Curie and Jacques Curie in 1880 while studying crystals of Quartz, Tourmaline, and Rochelle salt.\cite{boudia1997curie} After this discovery, numerous properties of materials have been studied extensively in relation to this phenomenon.
Some of the materials that exhibit piezoelectric properties are lithium niobate (LiNbO$_3$),\cite{yun2014lead} lead magnesium niobate-lead titanate (PMN-PT),\cite{hwang2014self} barium titanate (BaTiO$_3$),\cite{hu2016yue} zinc oxide (ZnO),\cite{ponnamma2019synthesis} lead zirconate titanate (PZT),\cite{jung2013energy,jain2015dielectric,qi2011enhanced} polyvinylidene fluoride (PVDF),\cite{tamang2015dna,bae2015characterization,hadimani2013continuous} etc.
Earlier studies suggested that lead zirconate titanate(PZT) has shown the highest piezoelectric response of -4.7 Cm$^{-2}$.\cite{cattan1999piezoelectric} But due to their rigidity, brittleness, and presence of toxic lead elements, their usage has been limited.\cite{Maeder2004} Therefore, numerous studies have been performed in search of lead-free materials that have better piezoelectric responses.\cite{park2020advances} Various 2D materials were synthesized experimentally, exhibiting ferroelectric effects.\cite{ wu2016intrinsic, wang2017two, fei2016ferroelectricity, mehboudi2016structural} These materials show spontaneous polarization below Curie temperature (T$_C$), for example in case of perovskite oxides, ferroelectric behavior originates because of a distorted crystal structure, which displaces the central atoms and eventually induces a permanent dipole moment within a crystal, due to which these types of materials are favorable for piezoelectric applications.\cite{megaw1952origin} In the recent research papers, Niobium oxide dihalides NbOX$_2$ (X= Cl, Br, and I) have been reported to be Van der Waals layered materials through computational screening, which show stable ferroelectric(FE) and anti-ferroelectric(AFE) phases in 3D as well as 2D layers. Moreover, polarization switching and structural phase transition were observed in NbOX$_2$ due to the co-existence of FE and AFE phases.\cite{ jia2019niobium} 

As reported by { Wu \it et al.} the monolayer of NbOI$_2$ displays excellent piezoelectric coefficients of about e$_{22}$= 32 $\times 10^{-10}$ Cm$^{-1}$ obtained using functional density perturbation theory.\cite{wu2022data} Moreover, it also shows a nonlinear response that can be tuned by applying strain, giving conversion efficiency of Second Harmonic Generation (SHG) of $> 0.2\% $\cite{ abdelwahab2022giant}. Because of the coexistence of the FE and AFE phases, the polarization can be switched easily.  NbOBr$_2$ and NbOCl$_2$ monolayers are also observed to exhibit a structural transition from paraelectric (PE) to the FE state by applying a small electric field. NbOBr$_2$ or NbOI$_2$ is suggested to be used as capacitors for energy storage, since their P-E loops show nearly zero hysteresis\cite{ jia2019niobium}. In this work, our objective is to examine the structural, electronic, elastic, mechanical, and piezoelectric properties of the bulk NbOX$_2$ (X = Cl, Br, and I) by using density functional theory (DFT). 

\section{Computational details}
Throughout this work, all calculations were performed within the density functional theory (DFT) framework, utilizing the PBE-GGA and hybrid PBE0 functionals, which rely on linear combinations of Gaussian-type functions (GTF) to describe the crystal orbitals, as implemented in the CRYSTAL17 code. The Nb, O, Cl, Br, and I atomic centers were described by the revised triple-$\zeta$ valence plus polarization (TZVP) basis set. During the structural optimization process, each compound was fully relaxed without imposing any constraints on the materials. The accuracy of convergence criteria was set at (10$^{-7}$, 10$^{-7}$, 10$^{-7}$, 10$^{-7}$, 10$^{-14}$). These parameters control the overlap and penetration for Coulomb integrals, the overlap for HF exchange integrals, and the pseudo-overlap, respectively. To ensure cell energy and structural parameters convergence, a criterion of $10^{-7}$ Ha was considered. The reciprocal space integration was carried out using a Monkhorst-Pack mesh of 10$ \times$ 10 $\times$ 10. Optimizations were controlled using analytical gradients with respect to atomic coordinates and unit-cell parameters, within a quasi-Newtonian scheme combined with Hessian Broyden-Fletcher Goldfarb-Shanno (BFGS) updating. For properties calculation, similar convergence criteria was considered but with a denser K-mesh of 12$\times$12$\times$12.

\section{Results and Discussion}
\subsection{Structural properties}
\begin{figure}[h!]     
	\centering
	\includegraphics[height=8cm,width=8cm]{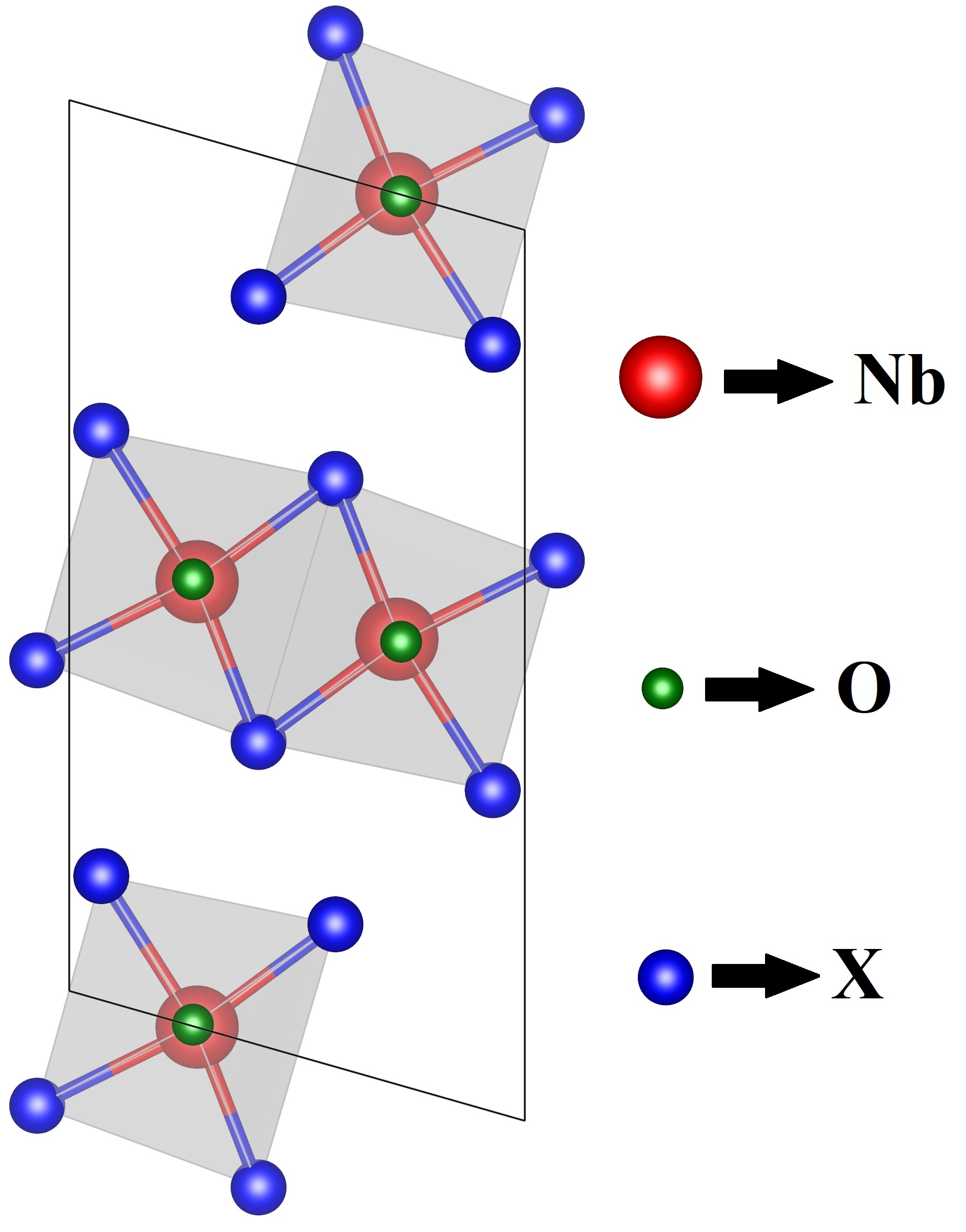} 
	\caption{Schematic representation of the bulk structure of NbOX$_2$(X=Cl, Br, and I), where red, green, and blue represent niobium, oxygen, and halide atoms respectively.}
	\label{fig:1}
\end{figure}

Bulk Niobium oxide dihalides NbOX$_2$ (X= Cl, Br, and I) belong to a group of metal oxide dihalides with a general chemical formula MOX$_2$, a layered Van der Waals material.\cite{rijnsdorp1978crystal,hillebrecht1997structural,beck2006crystal} They crystallizes into a monoclinic crystal having a space group C2.\cite{beck2006crystal} Whose structure is similar to perovskite oxides, as compared to other transition-metal oxide dihalides, NbOX$_2$ exhibits distorted octahedra in which Nb atoms are located at the center of an octahedral surrounded by 4 halides and 2 Oxygen atoms.\cite{tan2019two,zhang2021orbital,RIJNSDORP197879} Inside this, each octahedral niobium cations were displaced along the crystallographic a-axis, resulting in the formation of ferroelectric and antiferroelectric phases with broken inversion symmetry. Moreover, on the other planar axis, stable Peierls distortion among the Nb$^{4+}$ cations leads to the formation of dimers along the b-direction, which induces the crystal to exist in a stable ground state energy along with semiconducting and insulating properties. 
\cite{ jia2019niobium,fu2023manipulating,li2025peierls} Thus, the sheet is formed by joining two octahedra by sharing X-X edges along the c direction and O corners along the b direction. The sheet's surface contains halide atoms that form a square planar network. In the case of an adjacent sheet, the halide atoms lie over the center of these squares.\cite{ye2023manipulation} As shown in Fig.\ref{fig:1}, due to off-center displacement, there occurs a pairing distortion of Nb atoms in each NbOX$_2$ chain, which leads to the alternation of Nb-Nb atoms (that is, Nb-Nb...Nb) length along the b axis and unequal distances of Nb-O...Nb along a direction.\cite{whangbo1982importance} As reported by {Wahab \it et al.} in NbOI$_2$ system the Nb-Nb…Nb alternation is 3.17 and 4.35 {\AA}, in addition to Nb-O…Nb alternation is 1.82 and 2.10 {\AA} along the c- and b- directions respectively.\cite{abdelwahab2022giant}
Each van der Waals layer has been repeated in the a (vertical) direction. The exfoliation energy of NbOCl$_2$, NbOBr$_2$, and NbOI$_2$ was 0.22, 0.23, and 0.24 J$m^{-2}$, respectively as reported earlier, revealing their potential to be exfoliated into their corresponding monolayers from their bulk counterparts experimentally because of weak van der Waals interaction between each layer.\cite{mortazavi2022highly} In recent years NbOI$_2$ crystals were synthesized through self-vapor transport technology .\cite{fang20212d} To study its various properties theoretically, its structures were optimized using the functionals PBE-GGA and PBE0, whose optimized lattice parameters are given in Table\ref{1}. It is observed that our lattice parameters in the case of NbOI$_2$, are almost similar however NbOCl$_2$ and NbOBr$_2$ are nearly 1{\AA} less than the previously optimized structures.\cite{jia2019niobium,fang20212d} 
 
\begin{table*}[hbt!]
\small
\caption{Lattice parameters of the Bulk NbOX$_2$ (X=Cl, Br, and I)} 
\label{1}
\begin{tabular*}{\textwidth}{@{\extracolsep{\fill}}|l|l|l|l|l|l|l|l|l|l|}
\hline
X & \multicolumn{4}{c|}{PBE-GGA} & \multicolumn{4}{c|}{PBE0} & density$\rho$ (gCm$^{-3}$)\\
\hline
 & a(\AA) & b(\AA) & c(\AA) & volume(\AA$^3$) & a(\AA) & b(\AA) & c(\AA) & volume(\AA$^3$) &  \\
\hline
Cl & 13.404 & 3.885 & 6.763 &339.59 & 13.378 & 3.860 & 6.710 & 334.121 & 3.498\\
\hline
 Br & 14.578 & 3.888 & 7.024 &385.698 & 14.438 & 3.866 & 6.964 & 375.765 & 4.594\\
\hline
I & 15.764 & 3.879 & 7.538 & 444.113 & 15.418 & 3.861 & 7.471 & 428.152 & 5.425\\
\hline
\end{tabular*}
\end{table*}

\subsection{Electronic properties}
\begin{figure}[h!]     
	\centering
	\includegraphics[height=8cm,width=9cm]{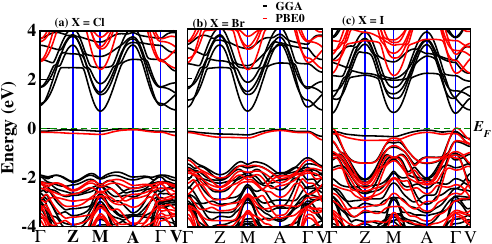}
	\caption{Band structure of bulk NbOX$_2$(X = Cl, Br, and I) obtained by using PBE-GGA (black) and PBEO (red) functional respectively. (a) NbOCl$_2$, (b) NbOBr$_2$, (c) NbOI$_2$.}
	\label{fig:2}
\end{figure}

To study the electronic properties of bulk NbOX$_2$ (X=Cl, Br, and I), the total density of states(DOS), the partial density of states (PDOS), and the band structure were calculated using the functionals PBE-GGA and PBE0 respectively. The bulk NbOX$_2$ in its ground state exhibits FE properties along with spontaneous polarization due to the Pseudo-Jahn Teller (PJT) effect, niobium ions were displaced along the niobium-oxygen bonds, distorting crystals along the a-axis, thus inducing the dipole moment at each pair of niobium-oxygen.\cite{jia2019niobium,tan2019two,ai2019intrinsic} Hence, these material shows phenomena like photovoltaic,\cite{rangel2017large,young2012first} Ferro-valley effects,\cite{wang2017two,shen2017electrically} and strong second harmonic generation due to the coupling between the FE ordering.
The band structure of bulk NbOX$_2$(X= Cl, Br, and I) is shown in Fig.\ref{fig:2} obtained by using the two functionals PBE-GGA and PBE0 respectively, these semiconductors exhibit an indirect band gap whose value is given in Table\ref{2}, it is observed that the band energy calculated using PBE0 gives better results than PBE-GGA compared to the value obtained experimentally. So, it indicates that the functional PBE-GGA underestimates the band energy values of the given materials because it fails to account for the vdW interactions.  The electronic band gap of NbOCl$_2$, NbOBr$_2$ and NbOI$_2$ bulk obtained with the PBE0 approach is 2.42 eV, 2.37 eV, and 2.13 eV, respectively, which shows a significant increment as compared to the previous report obtained by using the HSE06.\cite{mortazavi2022highly,jia2019niobium} However, the calculated band gap values are still less than the experimental values. 

\begin{figure*}[bth!]     
	\centering
	\includegraphics[height=10cm]{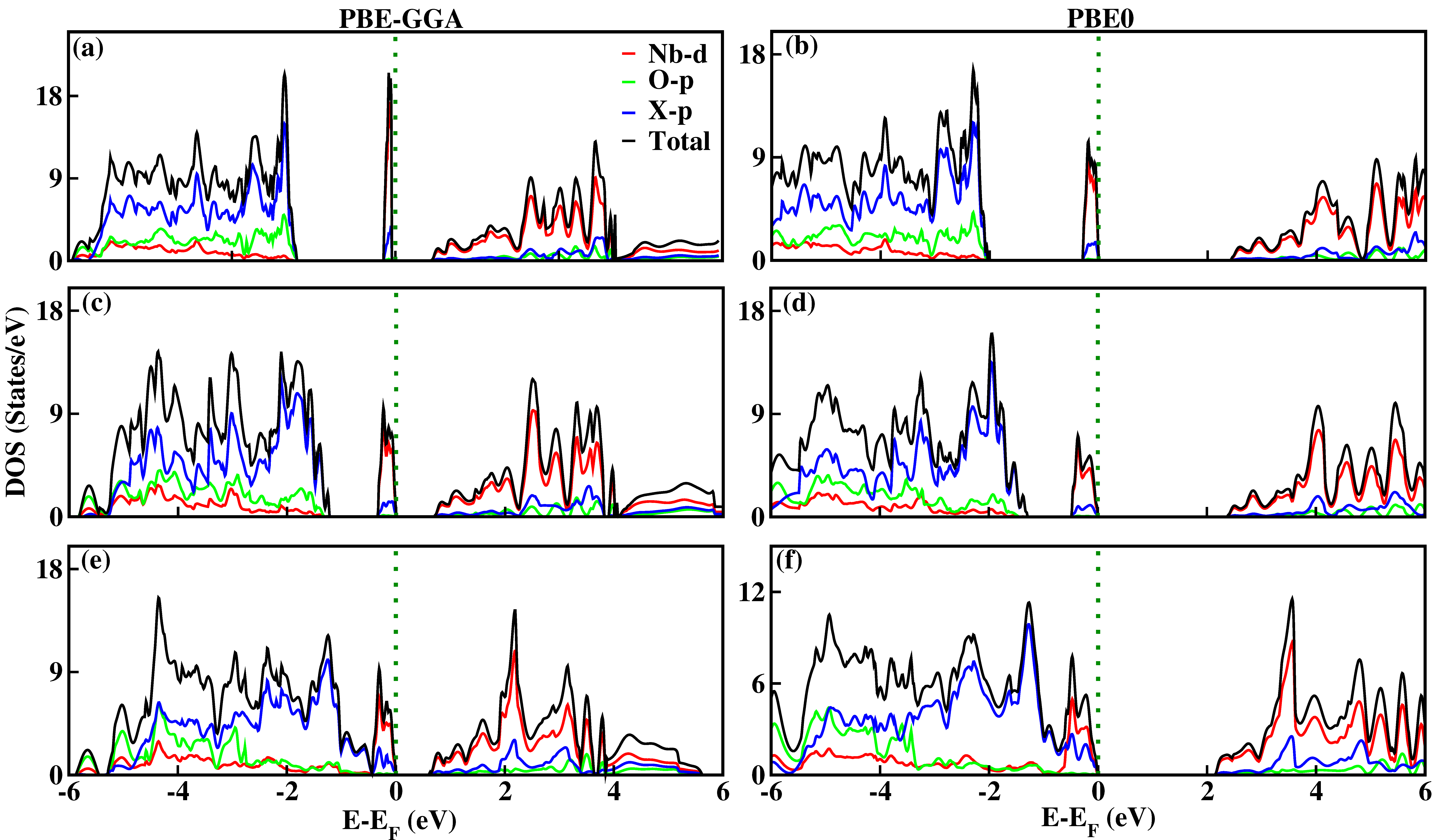}
	\caption{Left:(a, c, and e) are the partial density of states for NbOCl$_2$, NbOBr$_2$, and NbOI$_2$, respectively obtained from PBE-GGA. Right:(b, d, and f) represent the partial density of states for NbOCl$_2$, NbOBr$_2$, and NbOI$_2$, respectively, obtained by using PBE0. The Fermi level is taken as the reference point for plotting.}
	\label{fig:3}
\end{figure*}

From the density of states (DOS) plot of bulk NBOX$_2$(X= Cl, Br, and  I)
it is observed that the conduction band is mainly contributed by niobium atom d orbitals, and on the other hand, the valence band mainly consists of p orbitals from the halide and oxygen atoms. Moreover, there is an isolated energy band below the Fermi level of NbOCl$_2$ and NbOBr$_2$ contributed by niobium d orbitals, which is almost dispersionless throughout the Brillouin zone, as shown in Fig.\ref{fig:3}. A sharp peak in DOS is observed in compounds NbOCl$_2$ and NbOBr$_2$, which is mainly contributed by the Nb-$d_{z^2}$ and a small portion from the X-$p$ orbitals due to the dimerization of Nb atoms along the b-direction.\cite{jia2019niobium,guo2023ultrathin} 
Below the isolated electronic state, there is a gap between the rest of the valence bands, whose value is higher for Cl and decreases down the group as we reach I, and it merges with the other valence bands. This is mainly caused by an increase in the size of the atoms in the group that ultimately reduces the effect of dimerization as reported earlier.\cite{helmer2025computational} 
These materials have shown a unique property compared to other 2D materials, such as their electronic band structure is insensitive to the number of layers.\cite{guo2023ultrathin}

\begin{table}[htp]
    \centering
    \caption{Band gap energy of bulk NbOX$_2$(X = Cl, Br,and I) obtained using density functional PBE-GGA and PBE0 respectively}
    \small 
\begin{tabular}{|p{1.0cm}|p{1.50cm}|p{1.50cm}|p{1.5cm}|}
    \hline
	 \textbf{X} & \textbf{E$_{g}^{PBE0}$ (eV)} & \textbf{E$_{g}^{GGA}$ (eV)} & \textbf{E$_{g}^{Expt.}$ (eV)}  \\
		\hline 
Cl & 2.42  & 0.72  & 3.73\cite{jia2019niobium}  \\
        \hline
Br & 2.37  & 0.69  & 3.07\cite{beck2006crystal}  \\
        \hline
I & 2.13  & 0.60  & 1.83\cite{rijnsdorp1978crystal} \\
 \hline
\label{2}
\end{tabular}
\end {table}

\subsection{Elastic and mechanical properties}
Mechanical properties are the important characteristics of solids because they tell us their intrinsic ability to withstand any physical deformations imposed on them. Using finite strain theory\cite{weaver1976application} we compute the elastic constants of the bulk NbOX$_2$ materials.\cite{steddon1954finite} where elements of the elastic tensor is given by 
\begin{equation}
C_{ij} = \frac{1}{V}\frac{\partial^2E}{\partial\epsilon_i \partial\epsilon_j}
\end{equation}
where $\epsilon$ is the strain tensor, and i,j = 1,..,6(1 = xx, 2= yy, 3 = zz, 4 = yz, 5 = xz, 6 = xy).
Since the studied materials crystallize in a monoclinic symmetry must fulfill the following conditions for the system to be mechanically stable\cite{wu2007crystal} :
\begin{equation}
\begin{split}
C_{ii} > 0, \forall i \in \{1,6\},\\
C_{11} + C_{22} + C_{33} + 2(C_{12} +C_{13} + C_{23}) > 0 \\
C_{33}C_{55} - C^2_{35} > 0 \\
C_{44}C_{66} - C^2_{46} > 0 \\
C_{22} + C_{33} - 2C_{23} > 0 \\
C_{22}(C_{33}C_{55} - C^2_{35}) + \\2(C_{23}C_{25}C_{35} - C^2_{23}C_{55} - C^2_{25}C_{33}) > 0 \\
2[C_{15}C_{25}(C_{33}C_{12} - C_{13}C_{23})\\ + C_{15}C_{35}(C_{22}C_{13} - C_{12}C_{23}) +\\
C_{25}C_{35}(C_{11}C_{23} - C_{12}C_{13})] - \\
[C^2_{15}(C_{22}C_{33} - C^2_{13})\\
+ C^2_{35}(C_{11}C_{22} - C^2_{12})] + C_{55}(g) >0
\end{split}
\end{equation}

Where, g = $C_{11}C_{22}C_{33} - C_{11}C^2_{23} - C_{22}C^2_{13} - C_{33}C^2{12} + 2C_{12}C_{13}C_{23}$.\\
All the elastic constants of bulk NbOX$_2$ were observed to fulfill the Born stability criteria as given in Eqs.\ref{e1}.\cite{born1940stability} from the Table \ref{3} it is seen that elastic constants $C_{22}$ and $C_{33}$ are much higher than the $C_{11}$, $C_{44}$, $C_{55}$, and $C_{66}$ respectively. From Table\ref{4} it is observed that the values of Young's modulus fall under the range of Carbon fiber-reinforced plastic, implying good rigidity. Moreover, the compounds NbOCl$_2$ and NbOI$_2$ have higher values of Young's modulus(E) than NbOBr$_2$. All of the compounds show higher resistance to axial compression than shear deformations due to the presence of vdW interactions and exhibit positive eigenvalues.
The Kleinman coefficients ($\zeta$) calculated as\cite{kleinman1962deformation}
\begin{equation}
\zeta = \frac{C_{11} + 8C_{12}}{7C_{11} + 2C_{12}}
\label{e2}
\end{equation}
whose values range from $0 \le \zeta \ge 1$, which signifies bond bending and stretching when $\zeta \rightarrow 1$ and $\zeta \rightarrow 0$ respectively, from our calculation it seems that compounds under investigation(given in Table\ref{3}), NbOCl$_2$ and NbOBr$_2$ tend to have both bonds bending as well as stretching. However, on the other hand, NbOI$_2$ tends to have bond bending under stress.  
All the elastic constants calculated which are important for deriving the mechanical properties of the materials were represented in Table\ref{4} such as Bulk(K), Young’s (E), and Shear (G) moduli. 
From the value of elastic moduli, we will be able to calculate its brittleness and ductility. Using the following criteria we can distinguish the brittleness and ductile nature of the materials: below and above the critical value (0.26) of  Poisson’s ratio ($\nu$) indicates the brittle and ductile nature, similarly Pugh’s ratio (B/G) whose ratio below and above 1.75 signifies the brittle and ductile nature and as well as the Cauchy pressure given by (C$_{12}$-C$_{44}$), its positive(negative) values indicates ductile(brittle) nature of the materials respectively.
The data we obtained for the bulk NbOX$_2$ (X= Cl, Br, and I) for Poisson’s and Pugh’s ratios were 0.235 and 1.56, respectively below the critical points indicating a brittle nature and further confirmed by the negative values of Cauchy pressure.

\subsubsection{Elastic Anisotropic}
The Zener ratio $(A_{an})$ is used to check whether our studied materials are isotropic or anisotropic. For the material to be isotropic $A_n$ = 1, otherwise it is considered to be anisotropic. The value of $A_n$ is given by\cite{zhao2020crystal,shah2023dft}
\begin{equation}
 A_{an} = \frac{4C_{11}}{C_{11} + C_{33} - 2C_{13}}
 \label{e3}
\end{equation}
It is clear from the values obtained by using Eqs.\ref{e3} that the bulk NbOCl$_2$ is found to be nearly isotropic with 0.93. whereas, on the other hand NbOBr$_2$ is highly anisotropic followed by NbOI$_2$ as given in Table\ref{3}.
Therefore, the compounds under investigation followed the following trends with increasing anisotropy as NbOCl$_2$ < NbOI$_2$ < NbOBr$_2$.
In the case of lower symmetry crystal anisotropy depends on both the bulk as well as shear modulus, as observed from the table NbOBr$_2$ has the lowest values for bulk and shear modulus as compared to two other compounds. 

\subsubsection{Vicker’s Hardness}
Hardness is the fundamental property needed to describe the mechanical properties of solids fully, Vicker’s hardness $(V_H)$ is a semi-empirical formula employed to describe the hardness of materials depending on the  Young's modulus and Poisson's ratio which is given by\cite{dovale2022vickers} 
\begin{equation}
V_H = \frac{(1-2\nu)E}{6(1+\nu)}
\label{e4}
\end{equation}
Table\ref{3} shows the Vicker’s hardness values of the compounds under investigation, it is observed that NbOCl$_2$ and NbOI$_2$ exhibit higher values than NbOBr$_2$. However, all the materials under investigation are soft materials having Vickers hardness in the range (2-4) GPa, which is a little harder than the As$_2$Se$_3$ and GeAsSe.\cite{hartouni1986mechanical} Moreover, they were prone to mechanical wear under high stress or friction. Therefore, they need encapsulation or protective coatings when used in piezoelectric applications.

\subsubsection{Machinability Index}
It is a dimensionless quantity that helps manufacturers and engineers to select suitable materials and machining for specific applications which is described by the ratio of bulk modulus to shear resistance.\cite{ahmed2023dft}
\begin{equation}
\nu _m = \frac{K}{C_{44}}
\label{e5}
\end{equation}
The ratio greater than 1.45 is considered to be suitable for manufacturing. According to the machinability index, only NbOCl$_2$ and NbOI$_2$ are suitable for device fabrication as calculated using the above Eqs.\ref{e5}. However, NbOBr$_2$ has only 0.99 which is much lower than the limit value. 

\subsubsection{Melting Temperature}
For real-world applications, knowing the melting point of the materials is very crucial. To study the melting point of our studied materials the following equation is used\cite{chen2001calculation}
\begin{equation}
T_m = 553 + 5.911C_{11} \pm 300
\label{e6}
\end{equation}
All our studied materials show stability under high temperatures, out of which NbOCl$_2$ can withstand the temperature of about 710$\pm$ 300 K which is higher than the other compounds because chlorine is highly electronegative atoms resulting in the formation of stronger bonds than Br and I.

\subsubsection{Elastic wave velocities}
While studying the mechanical properties of the materials, elastic wave velocities are also one of the important parameters that can be estimated from the bulk and shear moduli and density of the material, which is given by Eqs. (\ref{e7} and \ref{e8}) \\
The longitudinal velocities,
\begin{equation}
\nu _l = \frac{[(B+\frac{4G}{3})]^{1/2}}{\rho}
\label{e7}
\end{equation}
And\\
Transverse velocities,
\begin{equation}
\nu _t = \frac{G}{\rho}^{1/2}
\label{e8}
\end{equation}
The above equations can be used to obtain mean elastic velocities\cite{rahman2024insight}
\begin{equation}
\nu _m = \frac{1}{3}{[\frac{1}{\nu _p ^3} + \frac{2}{\nu _s ^3}]}^{-1/3}
\label{e9}
\end{equation}
As represented in Table\ref{3} NbOCl$_2$ shows the highest translational and longitudinal velocities. However, NbOBr$_2$ and NbOI$_2$ show the highest mean elastic velocities, which means that they are stiffer than NbOCl$_2$.

\subsubsection{Debye Temperature{$(\theta _D)$}}
It is also an important parameter for designing and engineering new materials with particular properties. It generally gives the range of the temperature of a material contributing to its heat capacity. Higher Debye temperatures indicate stronger atomic bonds, which means stiffer and higher resistance to deformation. 
It is given by equation\ref{e10}
\begin{equation}
\theta _D = \frac{h\nu _D}{k}
\label{e10}
\end{equation}
Where, h is Planck’s constant, $\mu _D$ is Debye’s frequency, and k signifies Boltzmann constant.
To calculate the Debye temperature, we first need to find the Debye frequency $(\nu _D)$, which depends on the mean elastic velocities and the number of atoms per unit volume of the materials. The following equations represent Debye’s frequency\cite{chen2001calculation} as
\begin{equation}
\nu _D^3 = 2\pi 9\frac{N}{V}[\frac{2}{\nu _t^3} + \frac{1}{\nu _l^3}]^{-1}
\label{e11}
\end{equation}
using this equation it is found that NbO$Cl_2$ has the highest values of $\theta _D$ and $\nu _D$ indicating its thermally stable than NbOBr$_2$ and NbOI$_2$ respectively.

\begin{table*}[hbt!]
	\small
	\caption{Calculated elastic constant $C_{ij}$, Cauchy pressure ($C_{12}-C{44}$), Kleinman coefficient ($\zeta$), Anisotropic (A$_{an}$), Vicker's Hardness ($V_H$), Machinability Index ($\nu _m$), Melting Point (T$_m$), Sound Velocity ($\nu$), and Debye's Temperature ($\theta_D$). }
	\label{3}
	\begin{tabular*}{\textwidth}{@{\extracolsep{\fill}}|l|l|l|l|l|l|l|l|l|}
		\hline
X & $C_{11}$ (GPa) & $C_{12}$ (GPa) & $C_{13}$ (GPa) & $C_{15}$ (GPa) & $C_{22}$ (GPa) & $C_{23}$ (GPa) & $C_{25}$ (GPa) & $C_{33}$ (GPa)\\
 \hline
 Cl & 26.58 & 10.02 & 15.43 & 0.32 & 190.64 & 16.11 & -0.28  & 118.10\\
 
 Br & 8.80 & 3.51 & 4.56 & -2.03 & 154.32 & 15.52 & -1.95  & 105.08\\
 
 I & 11.30 & 7.91 & 7.13 & -0.07 & 330.54 & 6.19  & 0.56  & 89.07\\
 \hline
 X  & $C_{35}$ (GPa) & $C_{44}$ (GPa) & $C_{46}$ (GPa) & $C_{55}$ (GPa) & $C_{66}$ (GPa) & ($C_{12}-C_{44}$) (GPa) & $\zeta$ & A$_{an}$ \\
 \hline
 Cl  & -10.65 & 23.86 & -1.94 & 11.18 & 10.95 & -13.85 & 0.52 & 0.93\\
 
 Br  & -13.12 & 20.51 & -2.66 & 3.30 & 3.51 & -17.00 & 0.53 & 0.33\\
 
 I  & -10.12 &18.64 & -1.37 & 6.12 & 4.33 & -10.74 & 0.78 & 0.52\\
 \hline
 X & $V_H$ & $\nu _m$ & $T_m$ (K) & $V_l$ ($ms^{-1}$) & $V_t$ ($ms^{-1}$) & $V_m$ ($ms^{-1}$) & $\nu _D$  & $\theta _D$ (K)\\
 \hline
 Cl & 3.84 & 1.48 & 710.11 & 4309 & 2517 & 1250 & 16.81 & 807.12\\
 
 Br & 2.22 & 0.99 & 605.02 & 2856 & 1671 & 2000 & 10.76 & 516.83\\
 
 I & 3.68 & 1.71 & 619.79 & 3295 & 1936 & 2000 & 11.80 & 566.42\\
 \hline
	\end{tabular*}
\end{table*}

\begin{table*}[hbt!]
	\small
	\caption{Calculated values of elastic moduli: Bulk modulus, Young's modulus, Shear modulus, and Poisson's ratios$(\nu)$. The subscripts V, R, and H represent Voigt, Reuss, and Hill assumptions respectively.} 
	\label{4}
	\begin{tabular*}{\textwidth}{@{\extracolsep{\fill}}|l|l|l|l|l|l|l|}
		\hline
X & $K_V$ (GPa) & $K_R$ (GPa) & $K_H$ (GPa) & $E_V$ (GPa) & $E_R$ (GPa) & $E_H$ (GPa) \\
 \hline
Cl & 46.49 & 24.30 & 35.4 & 71.58 & 38.47 & 55.02 \\

Br & 35.04 & 5.69 & 20.37 & 54.11 & 9.49 & 31.80 \\

I & 52.60 & 11.00 & 31.8 & 82.14 & 18.46 & 50.31 \\
\hline
X & $G_V$ (GPa) & $G_R$ (GPa) & $G_H$ (GPa) & $\nu_v$ & $\nu_R$ & $\nu_H$\\
\hline
Cl & 28.78 & 15.56 & 22.17 & 0.24& 0.24 & 0.24\\

Br & 21.77 & 3.88 & 12.83 & 0.24 & 0.22 & 0.24\\

I & 33.13 & 7.56 & 20.34 & 0.24 & 0.22 & 0.23\\
\hline
	\end{tabular*}
\end{table*} 

\subsection{Piezoelectric properties}
Piezoelectric materials have a unique ability to convert mechanical stress into electrical energy. The conversion of mechanical stress into electrical energy depends on the amount of induced electric polarization inside the materials.\cite{cady1949theory,martin1972piezoelectricity} It was first discovered by the Curie brothers in 1880 in a quartz crystal,\cite{jaffe1958piezoelectric} after which there has been rapid technological development in the field of optoelectronic, information technology, high-resolution ultrasound devices and miniature filters for cellular communications and so on.\cite{ma2012phase,uchino1996piezoelectric,liu2018double}
In the absence of an external electric field, the total polarization of bulk materials is given by(P)\cite{C5TC04239A}: 
\begin{equation}
    P = P_p + P_{eq}
    \label{e12}
\end{equation} where $P_p$ is the polarization induced by strain, and $P_{eq}$ is the spontaneous polarization. In the case of CRYSTAL 17, it utilizes the Berry phase approximation to calculate the piezoelectric properties of a periodic solid. 
To study the piezoelectric responses of NbOX$_2$, it is important to calculate the direct ‘e’ and inverse 'd' piezoelectric tensors. These tensors give the changes in polarization ‘P’ generated by strain $‘\eta’$, which are induced by the application of stress and electric field (E) respectively. 
At constant stress,
\begin{equation}
\eta = d^T E
\end{equation}
At a constant electric field,
\begin{equation}
P = e\eta
\end{equation}
Polarization due to strain is measured in terms of intensity, and the Cartesian polarizations are calculated in terms of the strain tensors using the formula given below.
\begin{equation}
P_i = \displaystyle \sum _\nu e_{i\nu}\eta_\nu
\end{equation}
\begin{equation}
e_{i\nu} = \left(\frac{\delta P_i}{\delta\eta_\nu}\right) _E
\end{equation}
Here, i = x,y,z; $\nu$ = 1,2,3,4,5,6 (1 = xx, 2 = yy, 3 = zz, 4 = yz, 5 = xz, and 6 = xy); $\eta$ = strain tensor; and E = induced electric field. Here, the Berry phase (BP) approximation approach is used to calculate the polarization.\cite{aksel2010advances} The direct piezoelectric constants are presented n terms of the first numerical derivatives of the BP $\psi _l$.\cite{VANDERBILT2000147}
\begin{equation}
E_{i\nu} = \frac{|e|}{2piV} \displaystyle \sum _l a_{li}\frac{\delta \psi l}{\delta \eta _\nu}
\end{equation}
Where, a$_{li}$ is the i$^{th}$ Cartesian component of the l$^{th}$ direct lattice basis vector a$_l$. The derivative $\delta \psi$$_l$ is obtained computationally by applying finite strains to the lattice of the crystal structure. The relationship between the direct and converse piezoelectric tensors is given by
\begin{equation}
E = dC, 
\end{equation}
\begin{equation}
D = eS
\end{equation}
Where C denotes the fourth-rank elastic tensor and S = C$^{-1}$ denotes the fourth-rank compliance tensor.
Our calculated piezoelectric tensors are represented in Table\ref{5}, which are generated along the three axes x, y, and z with six different strains i.e. xx, yy, zz, yz, xz, and xy. Out of which the highest piezoelectric coefficients are observed along the y-axis with a positive yy-strain such as 4.99 Cm$^{-2}$, 4.693 Cm$^{-2}$, and 6.320 Cm$^{-2}$ for NbOCl$_2$, NbOBr$_2$, and NbOI$_2$ respectively. The different values of the piezoelectric coefficients obtained using different functionals are given in Table\ref{6}, especially for NbOI$_2$ piezoelectric coefficients are less by a small value (that is, 1.0 Cm$^{-2}$) while using the functionals PBE0 and PBE-D3. However, shows a large difference of almost 2.5 Cm$^{-2}$ compared to PBE0 and PBE. The dynamical charges in NbOI$_2$ have also been reported to be higher than the other two compounds studied because of the lower electro-negativity of iodine compared to chlorine and bromine. The smaller displacement off center in iodine contributes even more to piezoelectric responses\cite{posternak1994role} Among the bulk compounds studied, NbOI$_2$ exhibits the highest piezoelectric response even higher than PZT,\cite{cattan1999piezoelectric} and the recently reported value of Li-based perovskites LiSnCl$_3$ and LiSnBr$_3$.\cite{lalengmawia2024comprehensive}
Moreover, even the corresponding monolayer of NbOI$_2$ exhibits the highest piezoelectric response ($e_{22} = 31.6 \times 10^{-10}$ Cm$^{-1}$) reported by Wu {\it et al.}\cite{wu2022data}.

\begin{table}[hbt!]
	\small
	\caption{Piezoelectric tensor obtained using a PBE0 functional on a bulk NbOX$_2$(X= Cl, Br, and I) along the x, y, and z axes respectively all in.} 
	\label{5}
	\begin{tabular}{|p{0.6cm}|p{1.0cm}|p{1.6cm}|p{1.6cm}|p{1.6cm}|}
		\hline
\textbf{X} & \textbf{strains} & \textbf{x} (Cm$^{-2}$) & \textbf{y} (Cm$^{-2}$) & \textbf{z} (Cm$^{-2}$)\\
 \hline
Cl & xx & 0.00e-01 & -1.60e-02 & -0.00e-01 \\

 & yy & 0.00e-01 & 4.995 & 7.50e-02 \\

& zz & -0.00e-01 & -1.81e-01 & -3.00e-03 \\

& yz & -1.00e-03 & -3.00e-03 & 4.70e-02\\

& xz & -0.00e-01 & 1.40e-02 & -2.00e-03\\

& xy & 1.20e-02 & 1.00e-03 & -8.00e-03\\
\hline
Br & xx & 0.00e-01 & -6.00e-03 & 0.00e-01\\

& yy & 0.00e-01 & 4.64 & 0.00e-01\\

& zz & 0.00e-01 & -1.87e-01 & 0.00e-01\\

& yz & -4.00e-03 & 0.00e-01 & 4.20e-02\\

& xz & 0.00e-01 & 2.40e-02 & 0.00e-01\\

& xy & 1.2e-02 & 0.00e-01 & -6.00e-03\\
\hline
I & xx & 0.00e-01 & -1.3e-02 & 0.00e-01\\

& yy & 0.00e-01 & 6.32 & 0.00e-01\\

& zz & 0.00e-01 & -9.40e-02 & 0.00e-01\\

& yz & -6.00e-03 & 0.00e-01 & 5.60e-02\\

& xz & 0.00e-01 & 5.00e-03 & 0.00e-01\\

& xy & 1.00e-02 & 0.00e-01 & 1.00e-02\\
\hline
	\end{tabular}
\end{table} 

\begin{table}[hbt!]
	\small
	\caption{Piezoelectric responses obtained using a different functionals on a bulk NbOX$_2$(X= Cl, Br, and I).} 
	\label{6}
	\begin{tabular}{|p{1.6cm}|p{1.0cm}|p{1.6cm}|p{1.6cm}|}
		\hline
\textbf{Functionals} & \textbf{NbOCl$_2$} (Cm$^{-2}$) & \textbf{NbOBr$_2$} (Cm$^{-2}$) & \textbf{NbOI$_2$} (Cm$^{-2}$) \\
 \hline
 PBE0(our work) & 4.99 & 4.64 & 6.32\\
\hline
PBE\cite{wu2022data} & 3.6 & 3.4 & 3.8 \\
\hline
PBE-D3\cite{wu2022data} & 4.5 & 4.6 & 5.3 \\
\hline
	\end{tabular}
\end{table}

\section{Conclusion}
Here we study the structural, electronic, elastic, and mechanical properties of bulk NbOX$_2$(X=Cl, Br, and I) using the PBE0 functional incorporated in a DFT. Our calculation shows that all of the compounds studied are brittle along the axial direction. In the case of NbOI$_2$, due to the increased covalency, the bond tends to bend under the application of stress; however, in NbOCl$_2$ and NbOBr$_2$ both bond bending and stretching is possible. All of the compounds are anisotropic. NbOCl$_2$ is more thermally stable than NbOBr$_2$ and NbOI$_2$. The piezoelectric coefficients obtained for the bulk compounds NbOCl$_2$, NbOBr$_2,$ and NbOI$_2$ were 4.99 Cm$^{-2}$, 4.69 Cm$^{-2}$, and 6.32 Cm$^{-2}$ respectively and from the calculated machinability index indicates that NbOCl$_2$ and NbOI$_2$ can be used for device fabrication.

\begin{acknowledgments}
\textbf{DPR} acknowledges Anusandhan National Research Foundation (ANRF), Govt. of India, vide Sanction Order No.:CRG/2023/000310, \& dated:10 October, 2024.\\
A. Laref acknowledges support from the "Research Center of the Female Scientific and Medical Colleges",  Deanship of Scientific Research, King Saud University.\\
\end{acknowledgments}
\section*{Author contributions}
\textbf{L. D. Tamang:} Formal analysis, Visualization, Validation, Literature review, Performed Calculation, Writing-original draft, writing-review \& editing.\\
\textbf{L. Celestine:} Formal analysis, Visualisation, Validation, Literature review, Performed Calculation, writing-review \& editing.\\
\textbf{R. Zosiamliana:} Formal analysis, Visualization, Validation, writing-review \& editing. \\
\textbf{S. Gurung:} Formal analysis, Visualisation, Validation, writing-review \& editing. \\ 
\textbf{A. Laref}:Formal analysis, Visualisation, Validation, writing-review \& editing. \\	
\textbf{Shalika R. Bhandari:}Formal analysis, Visualization, Validation, writing-review \& editing. \\ 
\textbf{Tatyana Orlova:} Formal analysis, Visualization, Validation, writing-review \& editing. \\ 
\textbf{D. P. Rai:} Project management, Supervision, Resources, software, Formal analysis, Visualisation, Validation, writing-review \& editing. 

\section*{Data Availability Statement}


\begin{center}
\renewcommand\arraystretch{1.2}
\begin{tabular}{| >{\raggedright\arraybackslash}p{0.3\linewidth} | >{\raggedright\arraybackslash}p{0.65\linewidth} |}
\hline
\textbf{AVAILABILITY OF DATA} & \textbf{STATEMENT OF DATA AVAILABILITY}\\  
\hline
Data available on request from the authors
&
The data that support the findings of this study are available from the corresponding author upon reasonable request.
\\\hline
\end{tabular}
\end{center}

\nocite{*}

\bibliographystyle{aipnum4-1}
\bibliography{aipsamp}
\end{document}